\documentclass[3p,times,twocolumn]{elsarticle}
 \biboptions{comma,sort&compress}
 
\usepackage{graphicx}
\usepackage{here}
\usepackage{ecrc}

\volume{00}

\firstpage{1}

\journalname{Nuclear and Particle Physics Proceedings}

\runauth{}

\jid{nppp}

\jnltitlelogo{Nuclear and Particle Physics Proceedings}

\usepackage{amssymb}

\usepackage[figuresright]{rotating}

\usepackage{bm}
\usepackage{accents}

\newcommand{\SM}{\,{\rm SM}}

\newcommand{\diag}{{\textrm {diag}}}

\begin{document}

\begin{frontmatter}

\title{Flavor anomalies from warped space$^*$}
 \cortext[cor0]{Talk given at 20th International Conference in Quantum Chromodynamics (QCD 17),  3 - 7 july 2017, Montpellier - FR}
 \author[label1]{Eugenio~Meg\'{\i}as\fnref{fn1}}
 \fntext[fn1]{Speaker, Corresponding author.}
 \ead{eugenio.megias@ehu.eus}
 \address[label1]{Departamento de F\'{\i}sica Te\'orica, Universidad del Pa\'{\i}s Vasco UPV/EHU, Apartado 644, 48080 Bilbao, Spain}

 \author[label2
 ]{Mariano~Quir\'os}
 \ead{quiros@ifae.es}

 \author[label2]{Lindber~Salas}
 \ead{lsalas@ifae.es}

 \address[label2]{Institut de F\'{\i}sica d'Altes Energies (IFAE), and The Barcelona Institute of  Science and Technology (BIST), \\ Campus UAB, 08193 Bellaterra (Barcelona) Spain}


\pagestyle{myheadings}
\markright{ }
\begin{abstract}
We study the recently found anomalies in $B$-meson decays within a scenario with a warped extra dimension where the Standard Model (SM) fermions are propagating in the bulk. The anomalies are then interpreted as the result of the exchange of heavy vector resonances with electroweak (EW) quantum numbers. The model naturally leads to lepton-flavor universality (LFU) violation when different flavor fermions are differently localized along the extra dimension, signaling a different degree of compositeness in the dual holographic theory.
\end{abstract}

\end{frontmatter}

\section{Introduction}
\label{sec:intro}

Recent results found by the LHCb collaboration in $B$-meson decays seem to point towards the existence of new physics (NP) beyond the Standard Model (SM). This collaboration has determined the ratio for $\mathcal{B}(\bar{B} \to \bar{K}^{(\ast)} \ell \ell)$ with $(\ell=e,\mu)$, for muons over electrons, yielding a deviation of $\sim 2.6\sigma$ with respect to the SM prediction~\cite{Aaij:2014ora}. Similar anomalies have been found by the Babar, Belle and LHCb collaborations in the decay $\bar B \to D^{(\ast)} \ell^- \bar \nu_\ell$, leading to a combined deviation of $\sim 4\sigma$~\cite{Lees:2012xj,Aaij:2015yra,Hirose:2016wfn}. These phenomena suggest LFU violation in the processes $b \to s\ell\ell \, (\ell=e,\mu)$ and $b \to c \ell \bar \nu_\ell \, (\ell = e,\mu,\tau)$, respectively.

One may ask what are the natural theories whose direct detection is hidden from the actual experiments, but that can accommodate possible explanations of existing anomalies. These theories should solve, at the same time, some theoretical issues of the SM of particle physics, like the Higgs {\it Hierarchy Problem}. There are two main ultraviolet (UV) completions of the SM which can solve this problem: i) supersymmetry, and ii) extra dimensions. In this work we are focusing on the latter set of theories, one of the most popular realizations being the Randall-Sundrum (RS) model~\cite{Randall:1999vf}, characterized by a warped extra dimension, by which the Planck scale is warped down to the TeV scale. Simple modifications of the RS set-up with a strong deformation of the AdS metric towards the infrared (IR) have been proposed, which allow to keep under control the corrections to the EW precision parameters~\cite{Cabrer:2009we,Cabrer:2011fb,Carmona:2011ib,Megias:2015ory}. These models also allow the possibility that the SM is part of a nearly-conformal sector~\cite{Megias:2014iwa,Megias:2016jcw}. In this work we will study the LFU violation, and try to accommodate the recent data on flavor anomalies, within this scenario.

\section[Benchmark model]{Benchmark model}
\label{sec:Benchmark_model}


We consider a scenario analogous to the usual RS set-up~\cite{Randall:1999vf}. Theories with a warped geometry are characterized by a $5$D metric $ds^2 = e^{-2A(y)} \eta_{\mu\nu} dx^\mu dx^\nu + dy^2$, where $y$ is the extra dimension, and two branes located at the UV $(y=0)$ and IR $(y=y_1)$. The action of the model corresponds to a dilaton-gravity system
\begin{eqnarray}
&&\hspace{-1.3cm}S_\phi =  M^3 \int d^4xdy \sqrt{-g}\left(R-\frac{1}{2}(\partial_M \phi)^2-V(\phi)\right) \nonumber \\
&&\hspace{-0.5cm}- M^3 \sum_{\alpha} \int d^4x dy \sqrt{-g} 2\mathcal V^\alpha(\phi)\delta(y-y_\alpha) \,,
\end{eqnarray}
where ${\mathcal V}^\alpha \; (\alpha=0,1)$ are the UV and IR $4$D brane potentials at $(y(\phi_0),y(\phi_1))$ respectively, and $M$ is the $5$D Planck scale. The brane dynamics should fix $(\phi_0,\phi_1)$ to get $A(\phi_1) - A(\phi_0) \approx 35$ in order to solve the {\it Hierarchy Problem}, as this implies $M_{\textrm{\footnotesize Planck}} \simeq 10^{15} M_{\textrm{\footnotesize KK}}$. This theory is characterized by the superpotential
\begin{equation}
W(\phi) = 6k \left( 1 + e^{a\phi} \right)^b \,,
\end{equation}
where $a$ and $b$ are real parameters, and $k$ is related to the curvature along the extra dimension. The relation between the scalar potential and the superpotential is $V(\phi) = \frac{1}{2} [W^\prime(\phi)]^2 - \frac{1}{3} W(\phi)^2$. In the following we will consider $a=0.15$ and $b=2$, see e.g. Refs.~\cite{Megias:2017ove,Megias:2016bde}, and $M_{\rm KK}=2$ TeV. With this choice we have $k\simeq 1.6\times 10^{17}$~GeV, $L_1^{-1}\simeq 1.48\times 10^{18}$ GeV (where $L_1$ is the curvature radius at the IR location) and $M\simeq 1.84\times 10^{18}$ GeV,  such that $ML_1\simeq 1.24$ which guarantees perturbativity in the 5D gravitational theory.

\subsection{Gauge bosons}
\label{subsec:gauge_bosons}
In order to introduce the EW sector in the theory, the model can be extended with gauge bosons. The action of the model is then $S = S_\phi + S_5$ with~\cite{Cabrer:2011fb}
\begin{eqnarray}
&&\hspace{-1.5cm} S_5 = \!\! \int \!\! d^4x dy\sqrt{-g}\left(-\frac{1}{4} \vec W^{2}_{MN}-\frac{1}{4}B_{MN}^2-|D_M H|^2-V(H)\right) \,, \nonumber  \\
  && \label{eq:S5}
\end{eqnarray}
where we have defined the $5$D ${\rm SU}(2)_L\times {\rm U}(1)_Y$ gauge bosons $W^i_M(x,y)$, $B_M(x,y)$ with $i=1,2,3$ and $M=(\mu,5)$, and the Higgs field~$H(x,y)$. Then EW symmetry breaking is triggered on the IR brane.

The gauge fields can be decomposed in KK modes as~$X_\mu(x,y) = \sum_n X_\mu^n(x)\cdot f_X^{(n)}(y)/\sqrt{y_1}$, where $f_X^{(n)}$ is the wave function of the $n$-KK mode in the extra dimension. Each mode satisfies a Schr\"odinger-like equation with a mass $m_X^{(n)}$. In this kind of scenarios, $m_X^{(1)}$ and the EW scale~$v$ are linked by $m_X^{(1)} \lesssim 4\pi v$, so that the mass of the first KK mode must not exceed a few TeV to avoid the so-called Little Hierarchy Problem between both scales. In the following we will consider $m_{Z}^{(1)} \simeq m_{W}^{(1)} = M_{\rm KK}$.

\subsection{Fermions}
\label{subsec:fermions}

The model can be extended with fermions $f_{L,R}$ (quarks and leptons) which propagate in the bulk of the extra dimension. Then the SM fermions correspond to the zero mode wave functions
\begin{equation}
f_{L,R}(x,y)=\frac{e^{(2-c_{L,R})A(y)}}{\displaystyle\left(\int dy\, e^{A(1-2 c_{L,R})} \right)^{1/2}} f_{L,R}(x) \,, \label{eq:fermion_zero_mode}
\end{equation}
which are characterized by constants $c_{f_{L,R}}$ associated to a 5D Dirac mass: ${\cal L}_5 = M_{f_{L,R}}(y) \bar\Psi \Psi$ with $M_{f_{L,R}}(y) = \mp c_{f_{L,R}} W(\phi(y))$. The location of the zero modes depends on the value of $c_{f_{L,R}}$, so that when $c_{f_{L,R}} < 0.5$ $(c_{f_{L,R}} > 0.5)$ fermions are localized towards the IR (UV) brane, and they can be interpreted as partly composite (almost elementary) in the dual theory. Then the Yukawa interactions with the Higgs are induced by the Lagrangian
\begin{equation}
\mathcal L_Y=\hat Y_{ij}^u H \bar Q_L^i u_R^j +\hat Y_{ij}^d H \bar Q_L^i d_R^j +\hat Y_{ij}^e H \bar \ell_L^i e_R^j \,,
\end{equation}
with $i,j=1,2,3$ labeling the generations of fermions, and $\hat Y^{u,d,e}_{ij}$ being the 5D Yukawa couplings. 

\begin{figure}[htb]
\centering
 \includegraphics[width=0.45\textwidth]{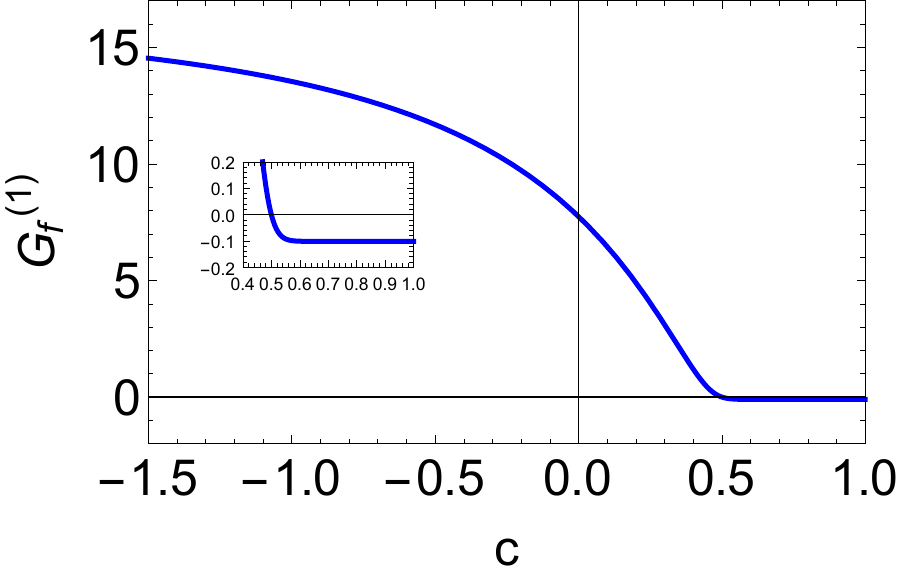} 
\vspace{-0.4cm}
 \caption{\it Coupling (normalized with respect to the 4D coupling) of a fermion zero-mode with the $n=1$ KK gauge field, $X_\mu^{n}$, as a function of the fermion localization parameter~$c$.
}
\label{fig:f1}
\end{figure}

A key ingredient of the model that will explain the anomalies, is the interaction of the KK modes of gauge bosons $(X = Z,\gamma)$ with leptons. This interaction reads
\begin{equation}
g^{\SM}_{f_{L,R}} \cdot G_{f}^{(n)}(c_{f_{L,R}}) \cdot X_\mu^n (x)\bar f_{L,R}(x)\gamma^\mu f_{L,R}(x) \,, 
\end{equation}
with $f = e, \mu, \tau$, and $G^{(n)}_{f}$ defined as
\begin{equation}
G^{(n)}_{f}=\frac{{\displaystyle \sqrt{y_1} \int e^{-3A} f_X^{(n)}(y) f^2(y)}}{\sqrt{\int [f_X^{(n)}(y)]^2 } \int e^{-3A}  f^2(y)  } \,,
\end{equation}
where $f(y)$ is the profile of the corresponding fermion zero-mode, as given by Eq.~(\ref{eq:fermion_zero_mode}). As one can see in Fig.~\ref{fig:f1}, 
fermions localized towards the IR (UV) brane interact strongly (weakly) with KK modes. As we will see below, LFU violation will be generated by a different degree of compositeness for different lepton flavors, i.e. different values of the parameters $c_{\ell_{L,R}}$ $(\ell = e, \mu, \tau)$.

\subsection{Electroweak observables}
\begin{figure}[htb]
\centering
 \includegraphics[width=0.48\textwidth]{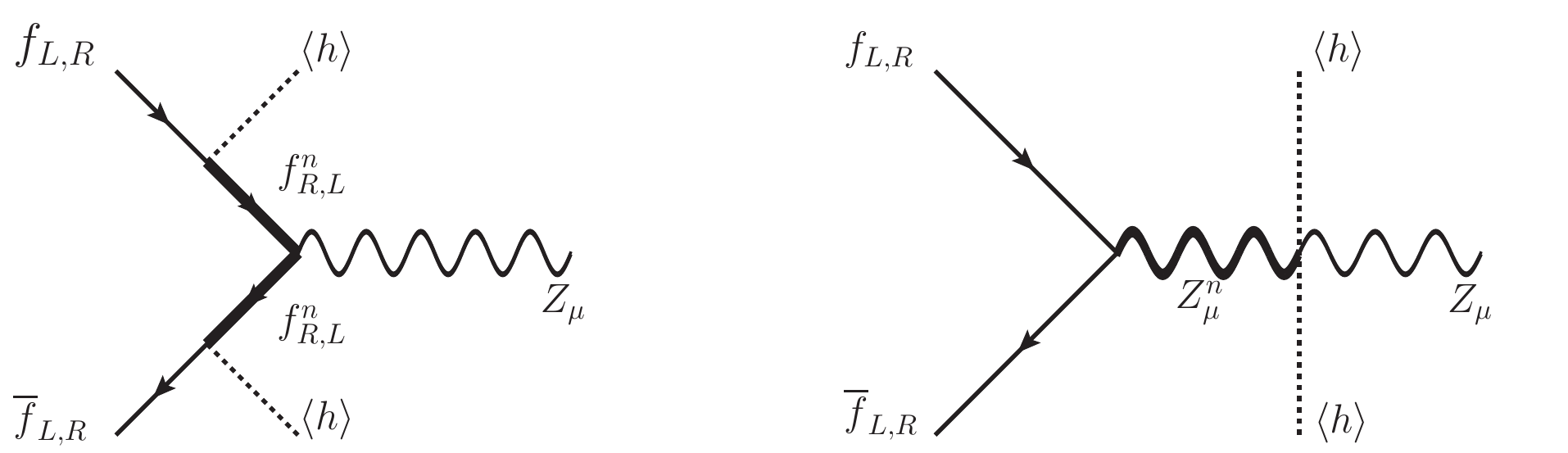} 
\vspace{-0.4cm}
 \caption{\it Diagrams contributing to $\delta g_{f_{L,R}} / g_{f_{L,R}}$.
  }
\label{fig:Zff}
\end{figure}

One of the most constraining observables is the $Z$ boson coupling to SM fermions $f_{L,R}$. This coupling is modified by two effects: i) contribution of vector KK modes, and ii) contribution from the fermion KK excitations. These effects are induced by the diagrams of Fig.~\ref{fig:Zff}.
When computing $\delta g_{f_{L,R}}$ for $f= e, \mu, \tau$ and~$b$, we get that the experimental bound $|\delta g_{f_{L,R}} / g_{f_{L,R}}| \lesssim 10^{-3}$~\cite{Olive:2016xmw} implies the mild constraint $c_{f_L} \gtrsim -0.5$. In addition, one can see from Fig.~\ref{fig:f1} that the coupling of EW and strong KK gauge bosons to a fermion~$f$ with $c_f = 0.5$ vanishes. Therefore if we assume $c_f \simeq 0.5$ for the first generation quarks $(f=u,d)$, it follows that Drell-Yan production of gauge bosons from light quarks is greatly suppressed.

\section{The $B$ anomalies}
\label{sec:LHCb_anomalies}

Flavor observables provide excellent probes for beyond SM physics. In particular $B$-meson decays due to the transitions $b \to s \ell^+ \ell^-$ and $b \to c \tau \bar\nu$ can be tested with a high accuracy at the LHCb, BaBar and Belle, leading recently to a significant deviation with respect to the SM predictions. In this section we will check if the explanation of this effect within the model presented above is consistent with all EW and flavor observables.

\subsection{The $\bar B\to \bar K^{(\ast)}\mu^+\mu^-$ anomaly}
\label{subsec:RK_anomaly}

LHCb measurements of branching ratios $\mathcal{B}(\bar{B} \to \bar{K} \ell \ell)$, $\ell = e , \mu$ lead to~\cite{Aaij:2014ora}
\begin{equation}
\hspace{-0.5cm} R_K\equiv R_K^{\mu/e}=\frac{\mathcal B(\bar B\to\bar K \mu\mu)}{\mathcal B(\bar B\to \bar K ee)}=0.745^{+0.090}_{-0.074}\pm 0.036 \,,
\end{equation}
which implies a deviation of $\sim 2.6\sigma$ with respect to the SM prediction $R_K^{\SM} = 1.0003(1)$~\cite{Bordone:2016gaq}. A similar result has been found recently for the ratio~$R_K^{\ast} \equiv \mathcal B(\bar B\to\bar K^{\ast} \mu\mu)/\mathcal B(\bar B\to \bar K^{\ast} ee)$~\cite{Aaij:2017vbb}. These anomalies can be interpreted in terms of $\Delta F = 1$ operators
\begin{equation}
\mathcal L_{eff}=\frac{G_F\alpha}{\sqrt{2}\,\pi} V^{\ast}_{ts}V_{tb}\sum_i C_i\,\mathcal O_i  \,,
\end{equation}
where the sum includes the semileptonic operators
\begin{eqnarray}
\mathcal O_9^{(\prime)\ell} &=(\bar s_{L(R)}\gamma_{\mu}  b_{L(R)})(\bar\ell \gamma^\mu\ell) \,, \nonumber\\
\mathcal O_{10}^{(\prime)\ell}&=(\bar s_{L(R)}\gamma_{\mu}  b_{L(R)})(\bar\ell \gamma^\mu\gamma_5\ell)\,, 
\end{eqnarray}
and Wilson coefficients $C_i = C_i^{\SM} + \Delta C_i$. Global fits to~$\Delta C_{9,10}^{(\prime) \mu}$ have been performed in the literature by using a set of observables concerning the $b\to s\ell\ell$ decay. From the recent fit of Refs.~\cite{Quim,Altmannshofer:2017fio}, we get the $2\sigma$ interval $\Delta C_9^\mu \in [-0.99,-0.38]$, where we have considered the approximate relation $\Delta C_9^\mu \simeq -\Delta C_{10}^\mu$ which naturally appears in our model.

\begin{figure}[htb]
\hspace{-0.2cm}    \includegraphics[width=0.50\textwidth]{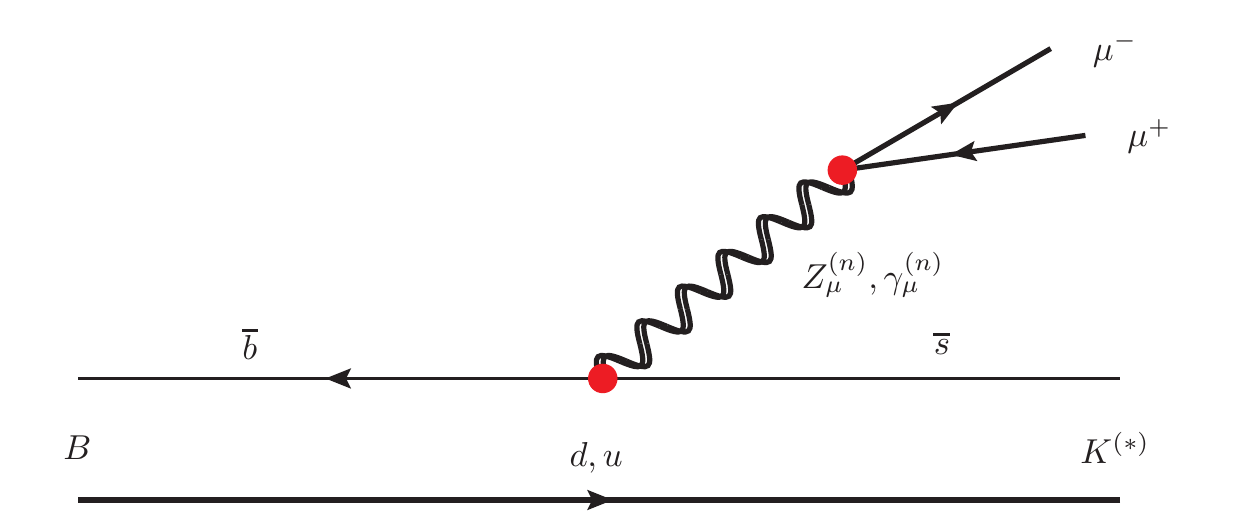} 
\vspace{-0.52cm}
 \caption{\it Diagram contributing to the $R_{K^{(\ast)}}$ anomalies.
  }
\label{fig:diagram}
\end{figure}

Within the model presented in Sec.~\ref{sec:Benchmark_model}, contact interactions can be obtained by the exchange of KK modes of the $Z$ boson and the photon as shown in Fig.~\ref{fig:diagram}, 
leading to the NP contributions to the Wilson coefficients
\begin{eqnarray}
\hspace{-1.3cm}&&\Delta C_9^{(\prime)\ell} = \! -(1-r)\!\sum_{X=Z,\gamma}\!\sum_n \frac{2\pi g^2 g_{\ell_V}^{X_n}\left(g^{X_n}_{b_{L(R)}}-g^{X_n}_{q_{L(R)}}\right)}{\sqrt{2} G_F\alpha c_W^2 M^2_n}\,, \\
\hspace{-1.3cm}&&\Delta C_{10}^{(\prime)\ell} = (1-r)\!\sum_{X=Z,\gamma}\!\sum_n \frac{2\pi g^2 g_{\ell_A}^{X_n}\left(g^{X_n}_{b_{L(R)}}-g^{X_n}_{q_{L(R)}}\right)}{\sqrt{2} G_F\alpha c_W^2 M^2_n}\,,
\label{C9}
\end{eqnarray}
with the couplings~$g_{f}^{Z_n}(c_{f}) = (T_{3f}-Q_f s^2_W)g G_{f}^{(n)}(c_{f})$, $g_{f}^{\gamma_n}(c_{f}) = Q_f s_W c_W g G_{f}^{(n)}(c_{f})$, and $s_W \equiv \sin \theta_W$, $c_W \equiv \cos \theta_W$. The unitary matrices $V_{u_{L,R}}$ and $V_{d_{L,R}}$ diagonalize the mass matrices for $u$ and $d$-type quarks. Their matrix elements, unlike those of the physical CKM matrix~$V \equiv V_{u_L}^\dagger V_{d_L}$,  are not measured experimentally. Given the hierarchical structure of the quark mass spectrum and mixing angles, in the following we will consider a Wolfenstein-like parametrization for the $V_{u_{L,R}}$ and $V_{d_{L,R}}$ matrices. Then the current $(\overline{b}\gamma^\mu s)$ comes from the entry $\left( V_d^\dagger G_d^{(n)} V_d \right)_{32}$ with $G_d^{(n)} = \diag(G^{(n)}_d(c_d), G^{(n)}_s(c_s), G_b^{(n)}(c_b))$.  We have defined the parameter $r = (V_{u_L}^*)_{32} (V_{d_L})_{33}/V_{cb}$.
\begin{figure*}[tbh]
\centering
 \begin{tabular}{c@{\hspace{4.5em}}c}
 \includegraphics[width=0.43\textwidth]{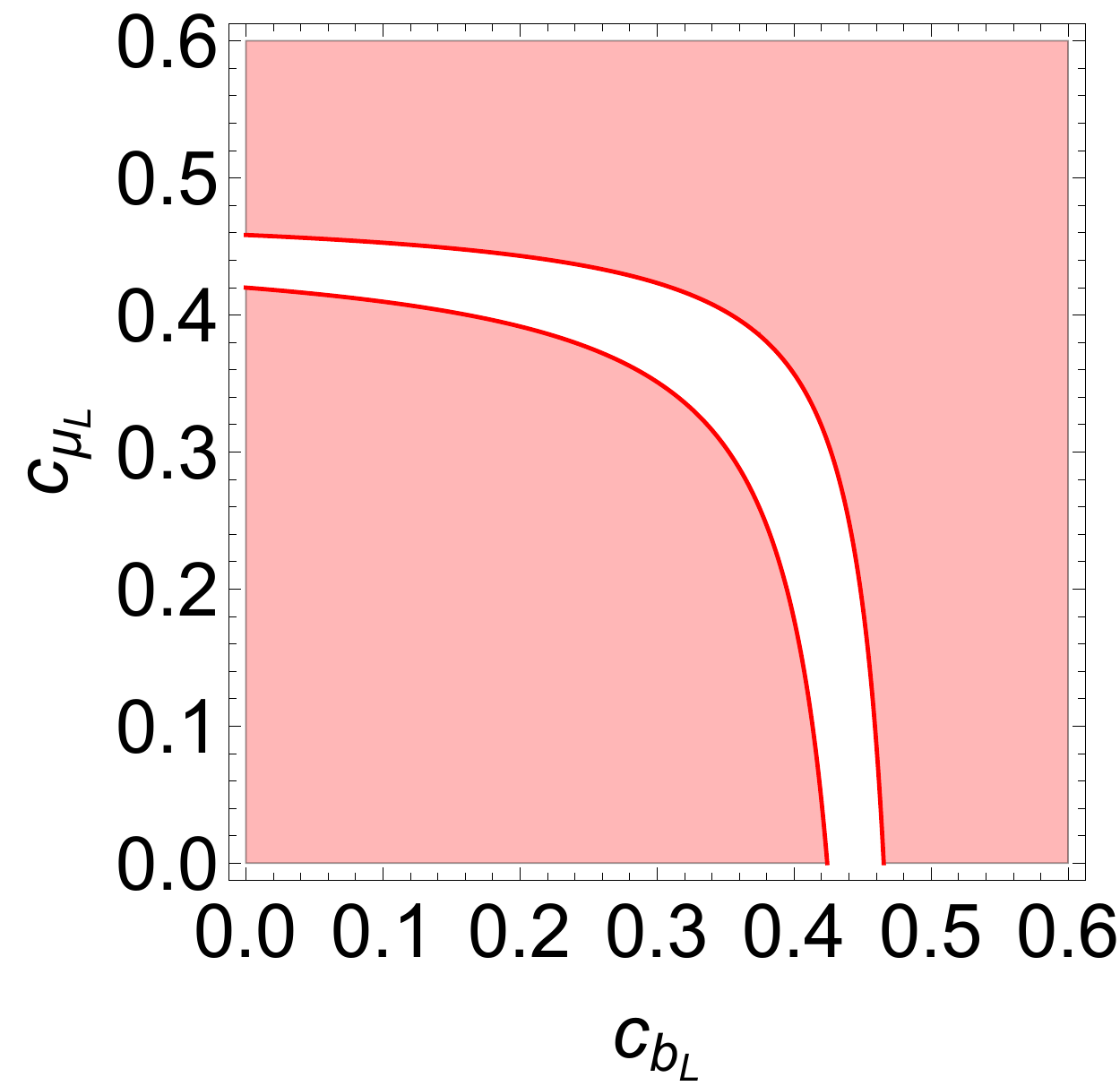} &
\includegraphics[width=0.443\textwidth]{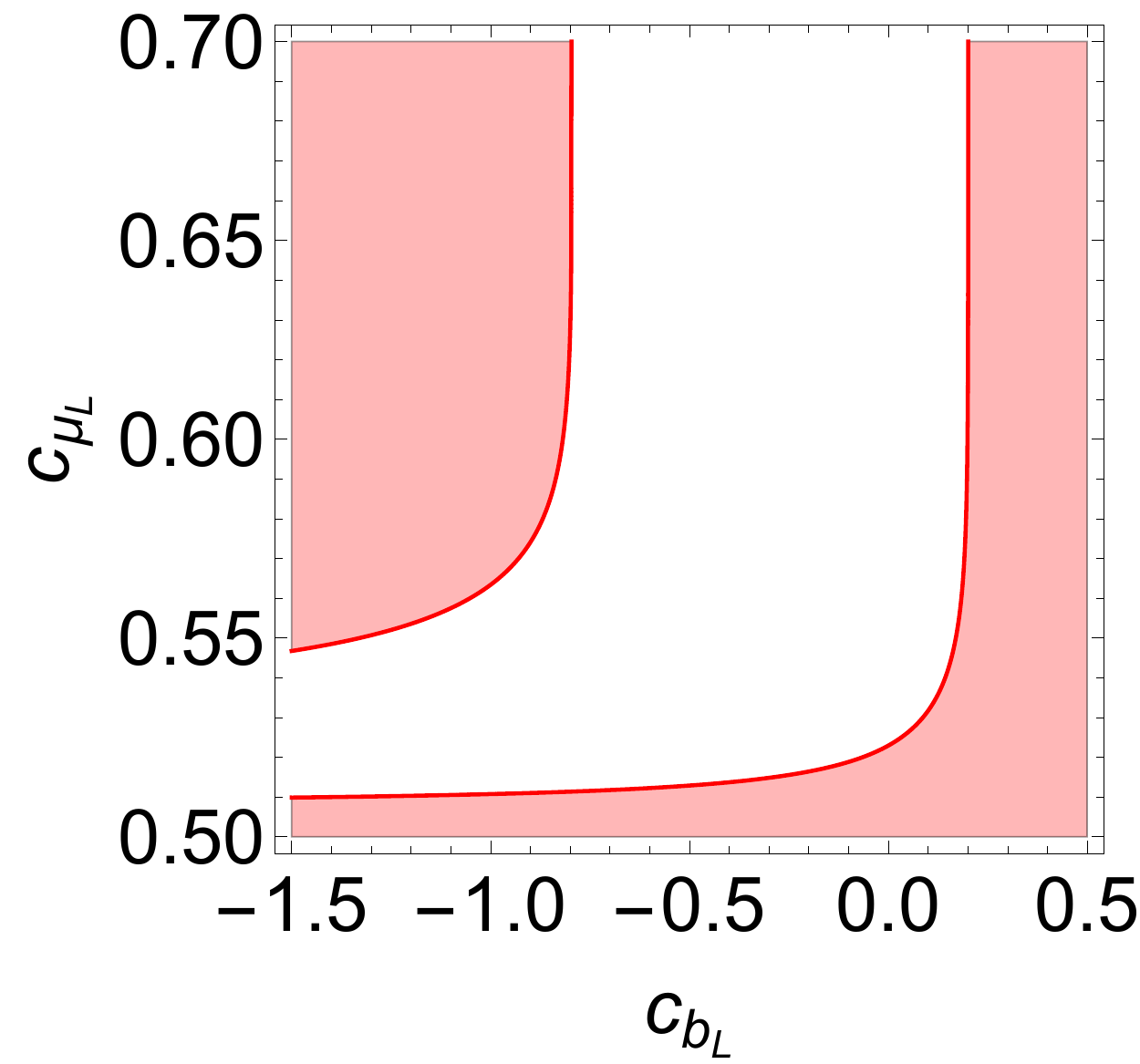} 
\end{tabular}
\vspace{-0.4cm}
 \caption{\it Region in the $(c_{b_L},c_{\mu_L})$ plane that accommodates the $2\sigma$ region $\Delta C_9^\mu \in [-0.99,-0.38]$ for $r=0.75$ (left panel) and $r=2.3$ (right panel).
  }
\label{fig:RK}
\end{figure*}
The region allowed by the global fits to the $\Delta C_9^{(\prime)\mu}$ Wilson coefficients is qualitatively different for $r<1$ and $r>1$ cases. We show both cases in Fig.~\ref{fig:RK} where the left (right) panel is for  $r=0.75$~\cite{Megias:2017ove} ($r=2.3$~\cite{Megias:2017vdg}) and we have fixed $c_{e_L}=0.5$. In both plots  the white region in the plane $(c_{b_L},c_{\mu_L})$ is allowed by data at the 95\% CL. As we can see from these plots $b_L$ is composite for both cases while $\mu_L$ is composite (elementary) for the $r<1$ ($r>1$) case.
%
%

A similar analysis can be done for the rare flavor-changing neutral current decay $B_s \to \mu^+ \mu^-$, which has been recently observed by the LHCb Collaboration with a branching fraction~$\mathcal{B}(B_s \to \mu^+ \mu^-) = (2.8^{+0.7}_{-0.6}) \times 10^{-9}$~\cite{CMS:2014xfa}. This measurement is quite consistent with the SM prediction~$\mathcal{B}(B_s \to \mu^+ \mu^-)_{\SM} = (3.66 \pm 0.23) \times 10^{-9}$~\cite{Bobeth:2013uxa}. The ratio $R_0 = \mathcal{B}(B_s \to \mu^+ \mu^-)/\mathcal{B}(B_s \to \mu^+ \mu^-)_{\SM} = 0.765^{+0.197}_{-0.171}$ leads at $1\sigma$ to an allowed region consistent with the one provided by~$R_{K^{(\ast)}}$. 

\subsection{The $R_{D^{(\ast)}}$ anomaly}
\label{sec:RD_anomaly}

\begin{figure}[htb]
\vspace{-0.3cm}\hspace{-0.2cm} \includegraphics[width=.50\textwidth]{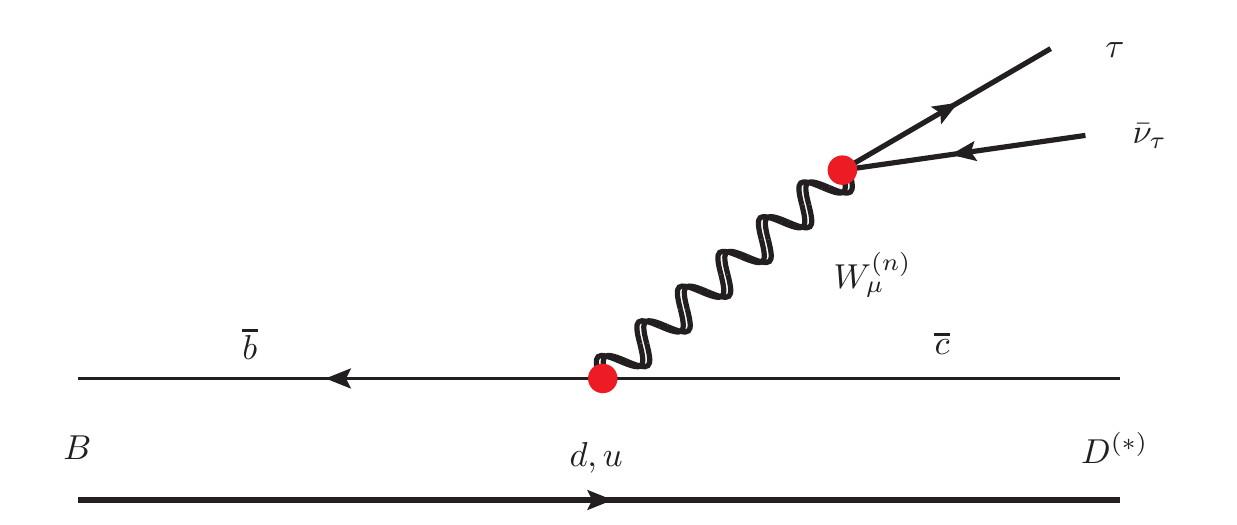}
\vspace{-0.52cm}
\caption{\it Diagram contributing to the $R_{D^{(\ast)}}$ anomalies.}
\label{fig:diagramRD}
\end{figure}

\begin{figure*}[tbh]
\centering
 \begin{tabular}{c@{\hspace{4.5em}}c}
 \includegraphics[width=0.43\textwidth]{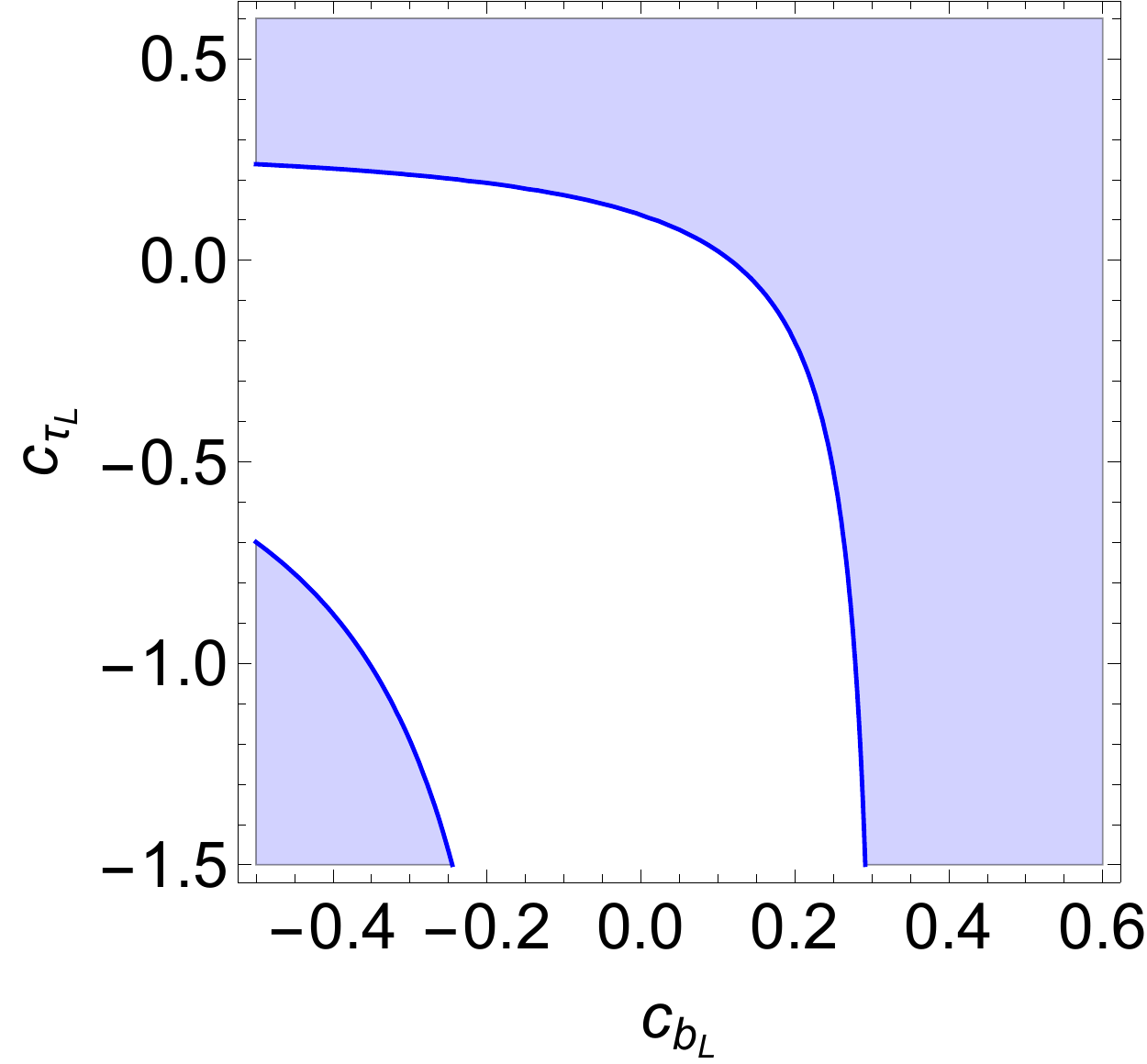} &
  \includegraphics[width=0.43\textwidth]{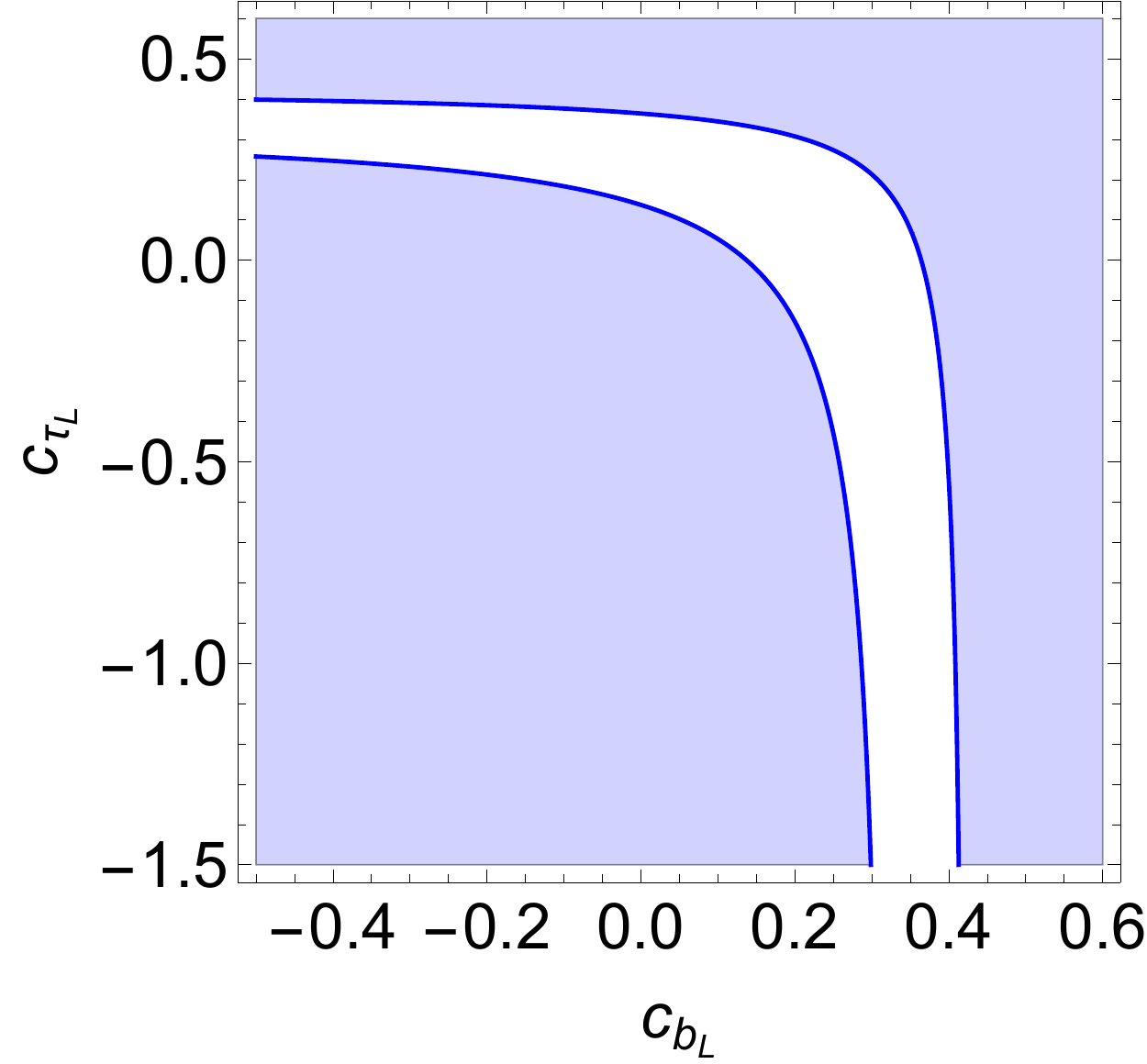} 
\end{tabular}
\vspace{-0.4cm}
 \caption{\it Region in the $(c_{b_L},c_{\tau_L})$ plane that accommodates the $2\sigma$ region of $(R_D,R_{D^{\ast}})$ observables for $r=0.75$ (left panel) and $r=2.3$ (right panel). We have considered $c_{\mu_L} = 0.44$ (left panel) and $c_{\mu_L} = 0.60$ (right panel).
  }
\label{fig:RD}
\end{figure*}

The $B$-meson decays due to the transition $b \to c \ell^- \bar \nu_\ell$ are being tested as well at $b$-factories and at the LHC, as they can be affected also by LFU violation. In particular the charged current decays $\bar B \to D^{(\ast)}\ell^-\bar\nu_\ell$ have been studied by the  BaBar~\cite{Lees:2012xj}, LHCb~\cite{Aaij:2015yra} and Belle~\cite{Hirose:2016wfn} collaborations. They measure the ratio
\begin{equation}
\hspace{-0.6cm}R_{D^{(\ast)}}\equiv R_{D^{(\ast)}}^{\tau/\ell}=\frac{\mathcal B(\bar B \to D^{(\ast)} \tau^- \bar \nu_\tau)}{\mathcal B(\bar B \to D^{(\ast)} \ell^- \bar \nu_\ell)} \quad (\ell = \mu \; \textrm{or} \; e) \,,
\end{equation}
with the experimental result~$R_D^{\rm exp}=0.403\pm 0.047$, $R_{D^{\ast}}^{\rm exp}=0.310\pm 0.017$, and a correlation~$\rho= -0.23$~\cite{Bernlochner:2017jka}. This result differs from the SM calculation $R_D^{\SM}=0.300\pm 0.011$, $R_{D^{\ast}}^{\SM}=0.254\pm 0.004$ by a deviation of $\sim 4\sigma$. Within the model of Sec.~\ref{sec:Benchmark_model}, the SM departure for $R_{D^{(\ast)}}$ is generated by the diagram of Fig.~\ref{fig:diagramRD}. 
In terms of the Wilson coefficients~$C^{\tau,\,\mu}= \sum_n (m_W^2/m_{W^{(n)}}^2) \cdot \left[ G_{q_L}^{(n)} + r(G_{b_L}^{(n)} - G_{q_L}^{(n)}) \right]G_{\tau_L,\,\mu_L}^{(n)}$, it reads~$R_{D^{(\ast)}}(C^\tau,C^\mu)=2 R_{D^{(\ast)}}^{\SM} \left|1+ C^\tau  \right|^2/(1+\left|1+ C^\mu  \right|^2)$. The regions allowed by $R_{D^{(\ast)}}$ data are displayed in Fig.~\ref{fig:RD} for $r=0.75$~\cite{Megias:2017ove} (left panel) and $r=2.3$~\cite{Megias:2017vdg} (right panel). We find that in both cases $b_L$ and $\tau_L$ fermions are localized towards the IR and thus show an important degree of compositeness although they are more composite for the case $r<1$.

\section{Constraints}
The main constraints are those from the experimental value of the coupling $g_{\tau_L}^Z$ and LFU tests, as e.g.~$\tau\to\mu\nu\bar\nu$ vs. $\mu\to e\nu\bar\nu$, as well as constraints from flavor physics.

\subsection{The coupling $Z\tau\bar\tau$}
The SM value of $g_{\tau_L}^Z$ receives tree-level corrections from KK modes of gauge bosons and fermions and leading loop corrections proportional to $Y_t^2$ as~\cite{Feruglio:2016gvd}
\begin{eqnarray}
\Delta g_{\ell_L}^Z&\simeq &\frac{v^2}{M_n^2}\frac{1}{16\pi^2}\left(3Y_t^2C^{t\ell}_n \log\frac{M_n}{m_t} +\mathcal O(g^4)\right) \,, \nonumber \\ 
\mathcal L_{eff}&=&\frac{C^{t\ell}_n}{M_n^2}  (\bar t_L\gamma_\mu t_L)(\ell_L \gamma^\mu \ell_L)+\dots \,,
\end{eqnarray}
while $g_{\tau_L}^Z=-0.26930(58)$~\cite{ALEPH:2005ab}, and $Y_t$ is the 4D top Yukawa coupling.

\subsection{Lepton universality tests}
The value of $R_{D^{(\ast)}}$ has also to agree with flavor universality tests in $\tau\to\mu\nu\bar\nu$  decays which are encoded~in
\begin{equation}
R_\tau^{\tau/\ell}=\frac{\mathcal B(\tau\to\ell\nu\bar\nu)/\mathcal B(\tau\to\ell\nu\bar\nu)_{\rm SM}}{\mathcal B(\mu\to e\nu\bar\nu)/\mathcal B(\mu\to e\nu\bar\nu)_{\rm SM}},
\end{equation}
with 95\% CL constraint $\ R_\tau^{\tau/\mu}\in [0.996,1.008]$~\cite{Pich:2013lsa,Feruglio:2016gvd}.
\subsection{Flavor observables}
\label{subsec:flavor}

New physics contributions to $\Delta F = 2$ processes come from the exchange of KK gluon modes. After integrating out the massive KK gluons, this gives rise to the following dimension six operator~\cite{Megias:2016bde,Megias:2017ove}
\begin{equation}
{\cal L}_{\Delta F = 2} = \sum_n \frac{c_{dij}^{LR(n)}}{M_n^2} (\overline d_{iR} d_{jL}) (\overline d_{iL} d_{jR}) \,, 
\end{equation}
\vspace{-0.1cm}
where 
\begin{equation}
c^{LR(n)}_{dij} = g_s^2 \prod_{\chi=L,R} \left[ (V_{d_\chi}^{\ast})_{3i} (V_{d_\chi})_{3j}\right] \left(G_{b_{\chi}}^{(n)} - G_{q_{\chi}}^{(n)}  \right) \,, 
\end{equation}
and similar expressions for operators $LL$ and $RR$, and for up quarks. The strongest current bounds for $d$-type quarks come from the operators $(\overline s_{L,R} \gamma^\mu d_{L,R})^2$ and $(\overline s_{R} d_{L})(\overline s_{L} d_{R})$, which contribute to the observables $\Delta m_K$ and $\epsilon_K$, while for $u$-type quarks they come from the operators $(\overline c_{L,R} \gamma^\mu u_{L,R})^2$ and $(\overline c_{R} u_{L})(\overline c_{L} u_{R})$, contributing to the observables $\Delta m_D$ and $\phi_D$~\cite{Isidori:2015oea}. 

\subsection{Combined results}
\label{subsec:final_result}

\begin{figure}[!htb]
\centering
      \includegraphics[width=.45\textwidth]{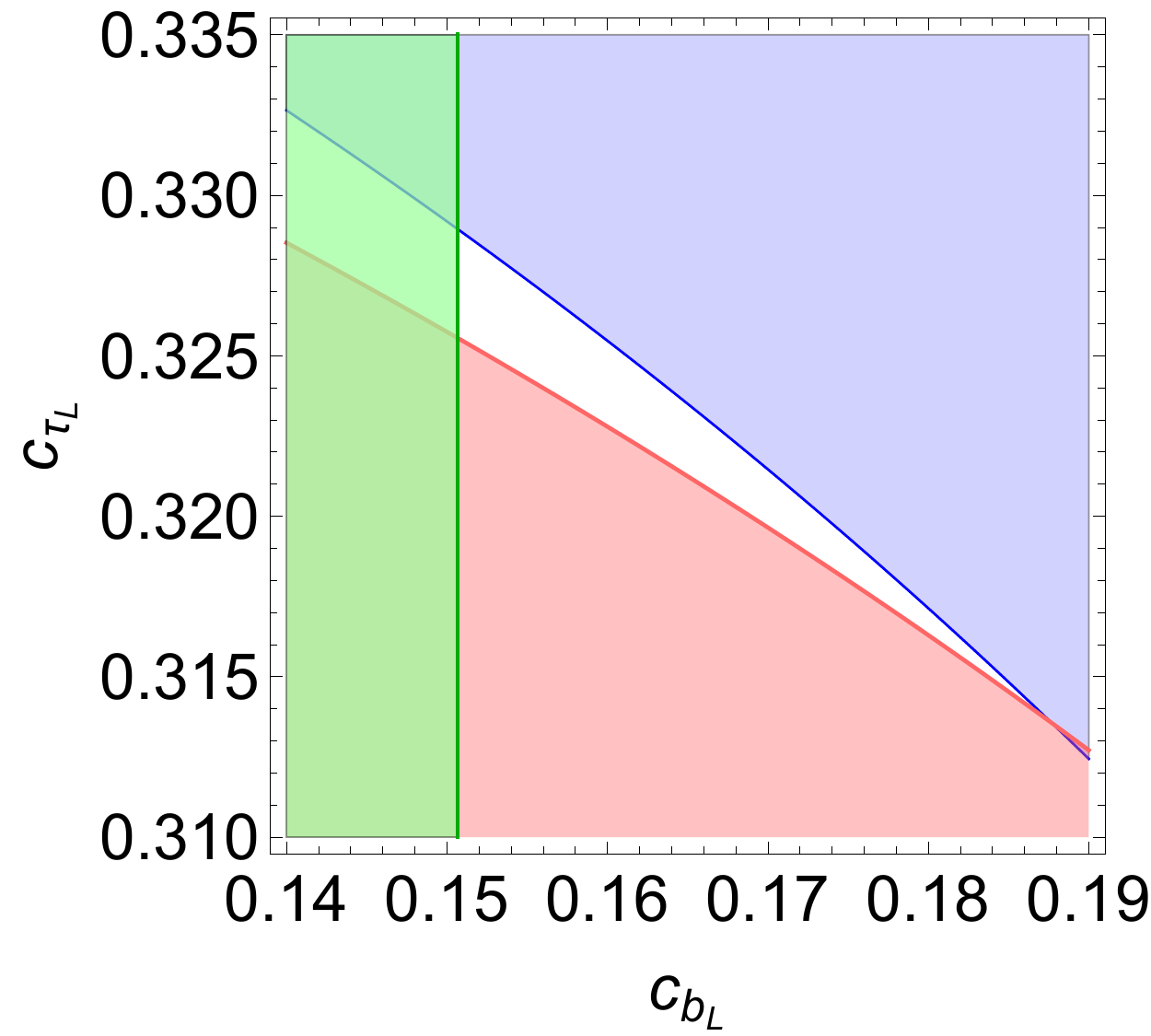}
\vspace{-0.4cm}
\caption{\it Region in the $(c_{b_L},c_{\mu_L})$ plane that accommodates all experimental constraints for $r=2.3$.}
\label{fig:combined}
\end{figure}

The previous constraints put very strong bounds on the available values of the parameters in the
plane $(c_{b_L},c_{\tau_L})$ as we show in Fig.~\ref{fig:combined}, where the blue region is 
forbidden by $R_{D^{(\ast)}}$, the red region is forbidden by $R^{\tau/\mu}_\tau$ and the green region is forbidden by the flavor constraints. The region excluded by $g^Z_{\tau_L}$ is outside the range of the plot.

\section{Conclusions}
\label{sec:conclusions}
\begin{figure}[!htb]
\centering
 \includegraphics[width=.4\textwidth]{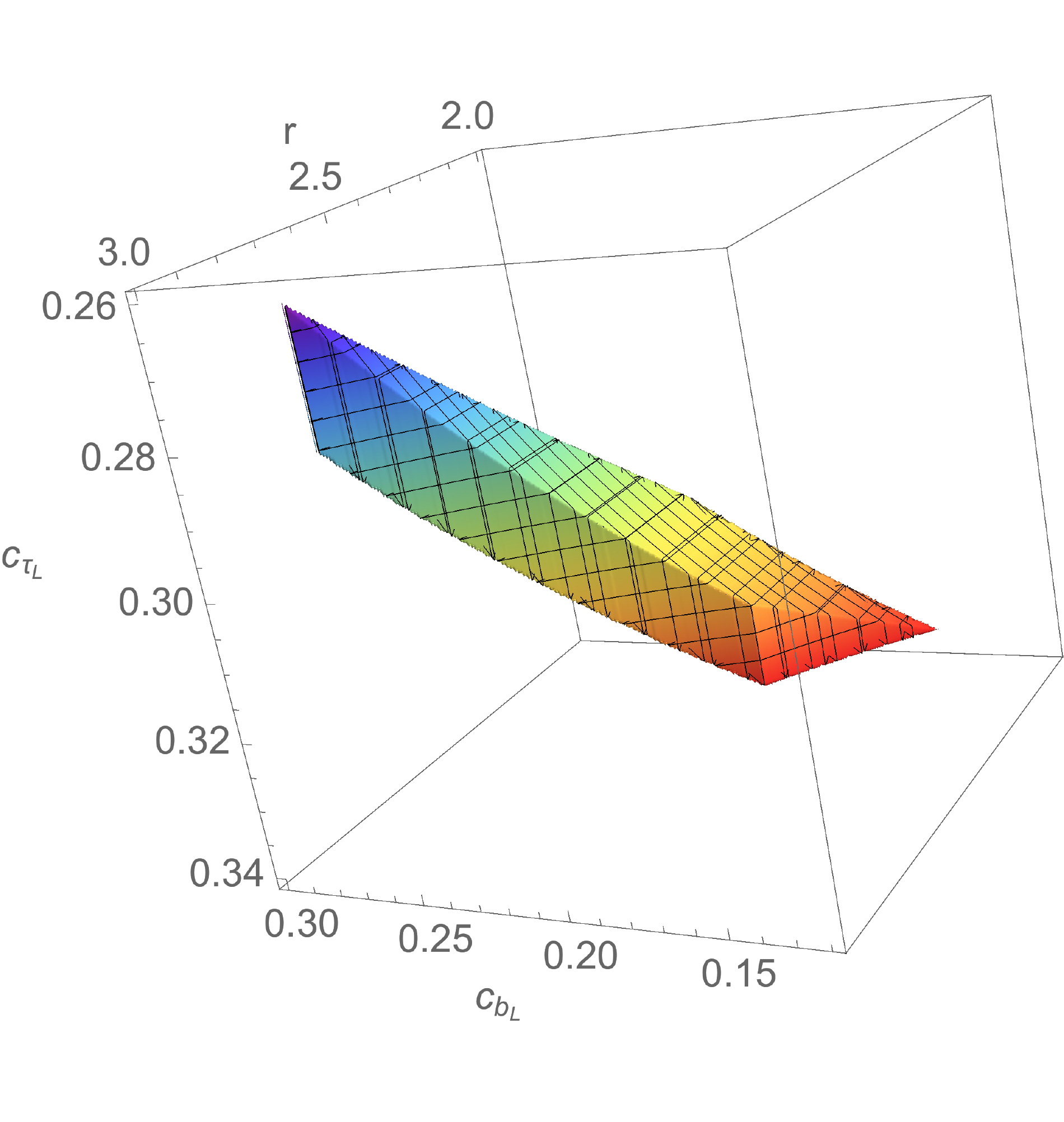}
\vspace{-0.4cm}
\caption{\it Region in the $(r,c_{b_L},c_{\tau_L})$ volume that accommodates all experimental data and electroweak and flavor constraints.
}
\label{fig:final}
\end{figure}

We have studied the compatibility of LFU violation data measured by the LHCb, BaBar and Belle collaborations, mainly in the observables $R_{K^{(\ast)}}$ and $R_{D^{(\ast)}}$ which appear to depart from the SM predictions, as well as all electroweak and flavor observables and universality tests as e.g.~$\tau\to\mu \nu\bar\nu$, in a model where the Higgs hierarchy problem is naturally solved by means of a warped extra dimension. We have found that the relevant variables are the set $(r,c_{b_L},c_{\tau_L})$, where the parameter $r$ was defined as $r = (V_{u_L}^*)_{32} (V_{d_L})_{33}/V_{cb}$, and $c_{b_L}$ and $c_{\tau_L}$ characterize the degree of compositeness of $b_L$ and $\tau_L$ respectively. We have seen in the previous sections that the values $r>1$ are preferred by experimental data, and we have presented some detailed analysis for $r=2.3$ where the third generation of fermion doublets, $b_L$ and $\tau_L$, is required to show a certain degree of compositeness. The general analysis in the volume $(r,c_{b_L},c_{\tau_L})$ is presented in Fig.~\ref{fig:final} where the allowed region at the 95\% CL 
is shown. The maximal region for the parameters is given by the range:
$
2.2<r<2.8, \ 0.14<c_{b_L}<0.28,  \ 0.265<c_{\tau_L}<0.33
$, while the first and second generation of quarks and leptons are elementary. These results are in good qualitative agreement with the hierarchy of quarks and charged lepton masses. 
The remaining LFU violation is the anomalous magnetic moment of the muon, which deviates from the SM by~$\sim 3.6\sigma$. In the context of warped theories, a solution was presented in Ref.~\cite{Megias:2017dzd} where heavy vector-like leptons were introduced.
Finally our model predicts, for any value of the parameters, the absolute ranges at 95\% CL for the branching ratios of $B\to  K^{(\ast)} \nu\bar\nu$ as~\cite{Megias:2017vdg}:
$1.14\ (2.70) \lesssim 10^5\mathcal B(B \to  K^{(\ast)} \nu\bar\nu)\lesssim 2.55\ (5.79)$,
%
as compared with experimental bounds (at 90\% CL) from Belle~\cite{Grygier:2017tzo} 
$10^5\mathcal B(B\to  K^{(\ast)} \nu\bar\nu)<1.6\ (2.7)
$, therefore on the verge of experimental discovery/exclusion!!

\section*{Acknowledgements} 
L.S. is supported by a \textit{Beca Predoctoral Severo Ochoa} of
Spanish MINEICO (SVP-2014-068850), and E.M. is supported by the
\textit{Universidad del Pa\'{\i}s Vasco} UPV/EHU, Bilbao, Spain, as a
Visiting Professor. The work of M.Q. and L.S.~is also partly supported
by Spanish MINEICO Grant FPA2014-55613-P, and by the
Severo Ochoa Excellence Program of MINEICO Grant
SO-2012-0234. The research of E.M. is also partly supported by Spanish
MINEICO Grant FPA2015-64041-C2-1-P, and by Basque Government Grant IT979-16.


\end{document}